\documentclass[preprint2]{aastex}

\begin{document}

\title{The filling factor - radius relation for 58 H~II regions across the disk of NGC~6946}


\author{Bernab\'e Cedr\'es \altaffilmark{1}, John E. Beckman \altaffilmark{1,2}, \'Angel Bongiovanni \altaffilmark{1}, Jordi Cepa \altaffilmark{1} and Andr\'es Asensio Ramos \altaffilmark{1}}
\affil{Instituto de Astrof\'{\i}sica de Canarias (IAC), E-38200 La Laguna, Tenerife, Spain}
\email{bce@iac.es}

\and

\author{Corrado Giammanco}
\affil{Dipartamento di Ingegneria Elettronica, Universit\`a di Roma Tor Vergata, Rome, Italy}

\and

\author{Antonio Cabrera-Lavers \altaffilmark{1,3}}
\affil{Instituto de Astrof\'{\i}sica de Canarias (IAC), E-38200 La Laguna, Tenerife, Spain}

\and
\author{Emilio J. Alfaro}
\affil{Instituto de Astrof\'{\i}sica de Andaluc\'{\i}a, CSIC, Glorieta de la Astronom\'{\i}a s/n, Apto. 3004, 18080 Granada, Spain}


\altaffiltext{1}{Departamento de Astrof\'{\i}sica, Universidad de La Laguna (ULL), E-38205 La Laguna, Tenerife, Spain}
\altaffiltext{2}{Consejo Superior de Investigaciones Cient\'{\i}ficas, Spain}
\altaffiltext{3}{Grantecan S. A., Centro de Astrof\'{\i}sica de La Palma, Cuesta de San Jos\'e, 38712, Bre\~na Baja, La Palma, Spain}

\begin{abstract}
Using the OSIRIS tunable narrow band imager on the 10.4m GTC (La Palma) we have mapped the SAB(rs)cd galaxy NGC 6946 over a $\sim$7.3x7.5 arcminute$^2$ field in the emission lines of the [SII]$\lambda\lambda$\, 6717, 6731 doublet, and in H$\alpha$. From these maps we have produced catalogs of the H$\alpha$ luminosities and effective radii of 557 HII regions across the disk, and derived the [SII] emission line ratios of 370 of these. The H$\alpha$ observations were used to derive the mean luminosity-weighted electron densities for the regions of the sample, while the [SII] line ratios allowed us to derive values of the {\it in situ} electron densities in the denser zones from which the major fraction of the radiation in these lines is emitted, for 58 of the regions. This is by far the largest data set of its kind for a single galaxy. A classical two phase model is  used to derive the filling factors of the regions. We find that although the mean electron density decreases with the square root of the radius of the regions, the {\it in situ} density is essentially independent of this radius. Thus the filling factor falls systematically, as the radius and the luminosity of the regions increases, with a power law of exponent -2.23 between filling factor and radius. These measurements should enhance the perspectives for more refined physical models of HII regions.
\end{abstract}


\section{Introduction}

The initial theory of the ionization equilibrium in H~II regions, proposed by Str\"omgren (1939), gave broad guidelines showing how their sizes and their emission line luminosities depend on the ionizing luminosities of their central stars and the density of the surrounding ISM.  It was not until twenty years later that Osterbrock and Flather (1959),  measured the electron density in the Orion Nebula using the ratio of the [OII] doublet lines $\lambda\lambda$3726,3729\,\AA\ and compared the result with a value of the rms electron density derived from the surface brightness in H$\beta$, finding a difference of an order of magnitude. They inferred that this H~II region must have a strongly inhomogeneous structure. Assuming that the [OII] line ratio is produced in a distributed fraction of the gas having higher electron density (what we call the {\it in situ} electron density), they suggested that the fractional volume occupied of these denser zones, the "filling factor" could be found by assuming a simple two phase model and comparing the two measured densities. 

This basic scenario has been used in many subsequent studies of H~II regions, both local and extragalactic. Simpson (1973) confirmed the deductions of Osterbrock and Flather (1959) for the Orion Nebula, and there have been many measurements of the {\it in situ} electron density in this, the nearest star forming zone, between then and the present day. We will cite here O'Dell \& Goss who combined 327.5\,MHz continuum observations with those of the optical [SII] doublet to explore spatial variations of the electron density and Rubin et al. (2011) who used the [SII] doublet and the [SIII] infrared doublet from Spitzer spectroscopy, to derive values of the local electron density in this region of a few hundred per cm$^{-3}$. It is worth mentioning a very different technique which yields the same qualitative result for HII regions. As an example Kassim et al.  (1989), using only radio continuum measurements at m and cm wavelengths also needed a highly inhomogeneous ISM to explain their measured intensity ratios. More generally the rms electron density can be derived from the radio continuum, and a value of the in situ density from recombination line ratios in the optical or near IR, with results for the latter which are canonically at least an order of magnitude greater than the former, pointing up the presence of the strong inhomogeneities in HII regions in general.

The filling factor of the denser zones of ionized interstellar regions is of interest if we want to quantify the fraction of ionizing photons which escape from the environments of OB star clusters, either into the diffuse gas of the galaxy, or even into the intergalactic medium. It is well understood that an inhomogeneous hydrogen cloud is more permeable to Lyc photons. Following the inference by Ferguson et al. (1996) that escaping Lyc from H~II regions in galaxies may be responsible for the ionization of the diffuse hydrogen well away from the regions, Zurita et al. (2002) produced a model in which the surface brightness distribution of the diffuse H$\alpha$ in NGC 157 could be reproduced in detail in a model with specified Lyc escape factors from the regions. Giammanco et al. (2004) showed how inhomogeneities, if optically thick, can alter the derivations of all parameters: electron temperature and density and their variances, ionization parameter, and abundance determinations. These considerations, not yet well enough studied, show the importance of quantifying density inhomogeneities. In consequence one would have expected to find in the literature many determinations of the key parameter: the filling factor. This is not, however, the case, basically because the derivation of the {\it in situ} densities using suitable emission line ratios is not trivially easy. Detailed measurements with the the [SII] doublet $\lambda\lambda$6717,6731 emission line ratio using long slit spectroscopy were published for 9 luminous H~II regions in 8 different galaxies by Casta\~neda et al (1992), but the most extensive data set is by Zaritsky, Kennicutt \& Huchra (1994) who measured 42 individual H~II regions in 39 galaxies. The range of those values, of order a few x 10$^2$ cm$^{-3}$, has been used repeatedly since then to make order of magnitude estimates of filling factors, by combining them with rms values of the electron density measured using the radius and H$\alpha$ surface brightness of  individual HII regions, (as outlined below in section 3.2).  Hunt \& Hirashita (2009) presented [SII] derived {\it in situ} electron densities, based on HST STIS data, for 24 HII regions, compact star forming zones, in a set of 24 galaxies, with corresponding rms densities obtained from H$\beta$ luminosities and radii, and derived the filling factors for all of these. However the observations of Casta\~neda et al. (1992), of Zaritsky et al. (1994) and of Hunt \& Hirashita (2009) do not permit a parametric study of behaviour within a single galaxy, to explore how the filling factor may vary systematically with e.g. the position of an HII region in the galaxy, or the overall physical parameters of the region itself: its H$\alpha$ luminosity and its radius.

In the present article we have used the capabilities of the OSIRIS tuneable narrow band filter on the 10.4m GTC (La Palma) to make narrow band images in the two emission lines of the [SII] doublet at 6716/6731\,\AA\ across the full face of the SAB(rs)cd type galaxy NGC 6946, from which we can derive quite a wide range of in situ electron densities for individual HII regions with a range of luminosities and radii. At the same time we obtained flux calibrated images in H$\alpha$, from which we could derive values of the rms electron density. For those regions with reliably determined values of both of these parameters we could find their filling factors, and examine how those parameters vary according to the radius and galactocentric radius of the regions.

\section{Observations and data processing}

The data employed in this study were obtained in an observing run during 2011 August 21 and 23, at the 10.4\,m Gran Telescopio Canarias (GTC), located at Roque de los Muchachos Observatory. The instrument used was OSIRIS, in tunable filter imaging mode, with a FOV of 7.3$^\prime\times$7.5$^\prime$. The mean seeing during the observations was about 0.7$^{\prime\prime}$.
We took a scan composed of 15 slices for H$\alpha$ and the [NII]$\lambda\lambda$6548,6584 lines, tuned to wavelengths from 6549\,\AA\ to 6633\,\AA\  in the optical centre, with FWHMs from 13\,\AA\  to 14\,\AA. Each slice consisted of 6 exposures of 35 seconds, three on target and three off-target in order to sample the sky.
For the [SII]$\lambda\lambda$6717,6731 lines, we took a scan composed of 19 slices, tuned to wavelengths from 6704\,\AA\ to 6812\,\AA\  in the optical center, with FWHM$\simeq$13\,\AA. Also, for these observations, each wavelength had 6 exposures of 90 seconds, three on target and three off-target.
In a forthcoming paper, we will discuss in detail the observation strategy and the reduction and calibration processes.
As a short summary, the data were reduced and calibrated using the IRAF package\footnote {IRAF is distributed by the National Optical Astronomy Observatory, which is operated by the Association of Universities for Research in Astronomy (AURA) under cooperative agreement with the National Science Foundation.}. Each image at each wavelength was bias-subtracted using bias frames taken during the night. The images for each wavelength were then flatfield-corrected using master flats obtained by the median combination of dome flatfields (five for each wavelength) and then corrected from ilumination by multiplying them by a normalized surface obtained from a twilight sky r$^\prime$ flatfield. 
Sky images for each wavelength were created by combining (with the median) the 3 off-target images after masking the stars. The sky image corresponding to a given wavelength was subtracted from the on-target images.
After that, all the on-target images were aligned and an astrometrical solution for each one was calculated. The position of the optical axis for each on-target image at each wavelength was determined using the positions of several image-ghost pairs.
Employing the program {\tt SExtractor} (Bertin \& Arnouts, 1996) and the sum of all H$\alpha$ wavelengths as the detection image, a catalogue of H~II regions for each wavelength was created. A region was considered detected when it presented a flux per pixel 3 sigma over the background, and when the total area (defined as the isophotal area) was larger than the area subtended by a circle with radius equal to the seeing of the image. The equivalent radius, $r$, for each H~II region is defined as the square root of the isophotal area of the region divided by $\pi$.

For each H~II region detected in each tuned slice, the corresponding real wavelength $\lambda$ was calculated, following Gonz\'alez et al. (2013, in preparation\footnote {For more information, see the webpage http://gtc-osiris.blogspot.com.es/2012/03/new-rtf-calibration.html and follow the links there.}), as
\begin{equation}
\lambda=\lambda_0-5.04r_o^2+a_3(\lambda)r_o^3.
\end{equation}
Where $\lambda_0$ is the central wavelength of a given slice, $r_o$ is the distance of the H~II region to the optical centre in arcmin, and $a_3(\lambda)$ is defined as
\begin{equation}
a_3=6.17808-1.6024\times10^{-3}\lambda+1.0215\times10^{-7}\lambda^2
\end{equation}
  
For each region detected, the pseudospectra of the H$\alpha$ zone and the [SII] zone were generated. The data were then flux calibrated following the prescription by Jones et al. (2002) for an equivalent system to that used here.

In order to obtain the amplitudes of the expected components, we performed a least squares inversion of the spectral profile. The method uses a combination of Gaussians whose number and widths are pre-assigned according to the expected emission lines, leaving their amplitude and central wavelengths to vary freely inside a range. The Gaussians are then convolved with the spectral response function of the instrument (an Airy function, see Jones et al. 2002); solving the least-squares problem yields the amplitudes and central wavelengths of all the components.

The errors for each line were calculated as the addition of the uncertainty in the measurement of the total integrated flux of the region and the uncertainty in the calibration. The error in the fitting process is small enough, when compared with the previous ones, to be considered negligible for the calculation. The mean errors in the line intensities are 5\% for H$\alpha$ and 7\% for the [SII] doublet.

\begin{figure*}[!p]
 \centering
 \begin{tabular}{c c}
  \multicolumn{2}{c}{\includegraphics[width=16cm]{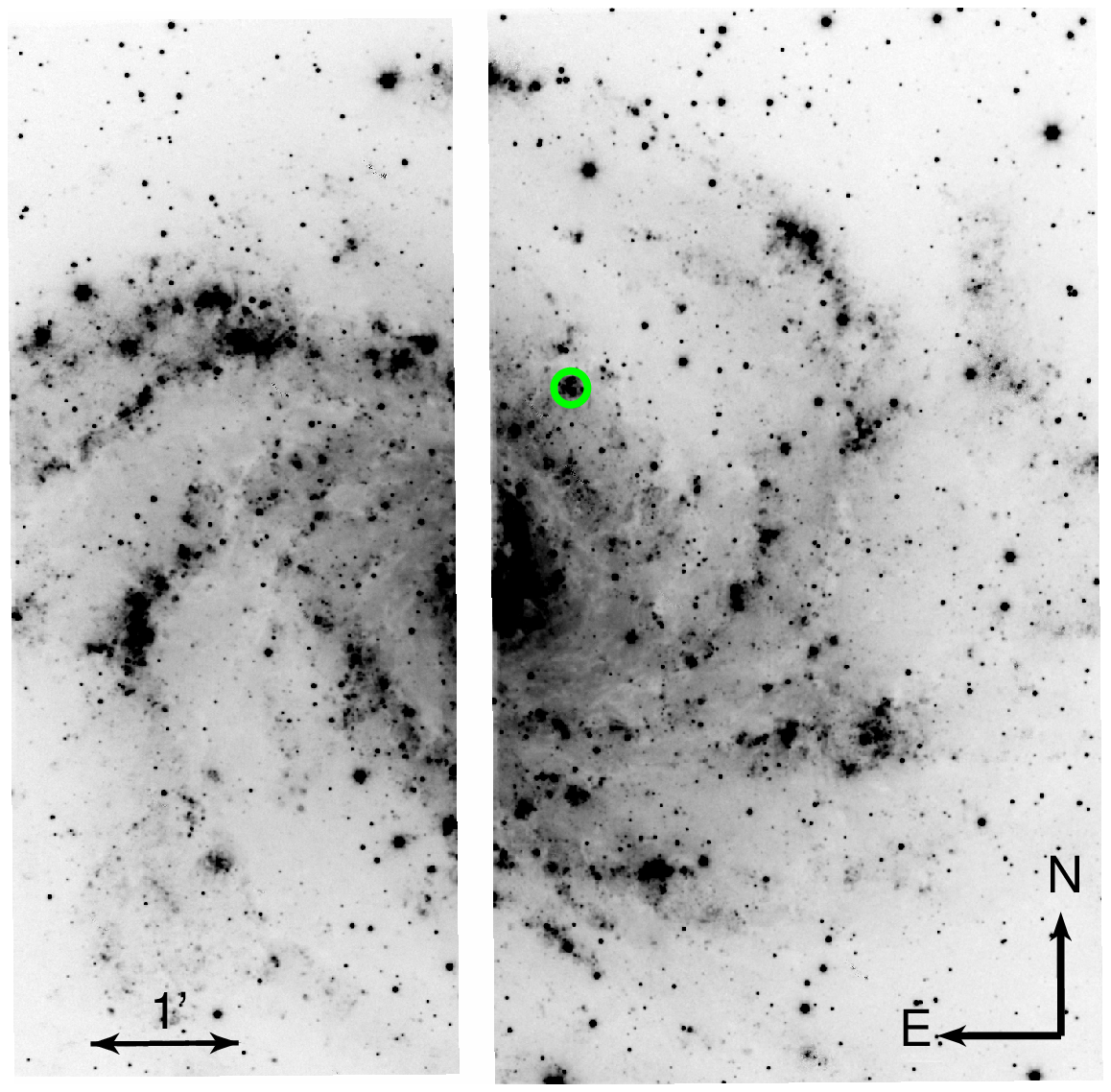}}\\
  \includegraphics[width=8cm]{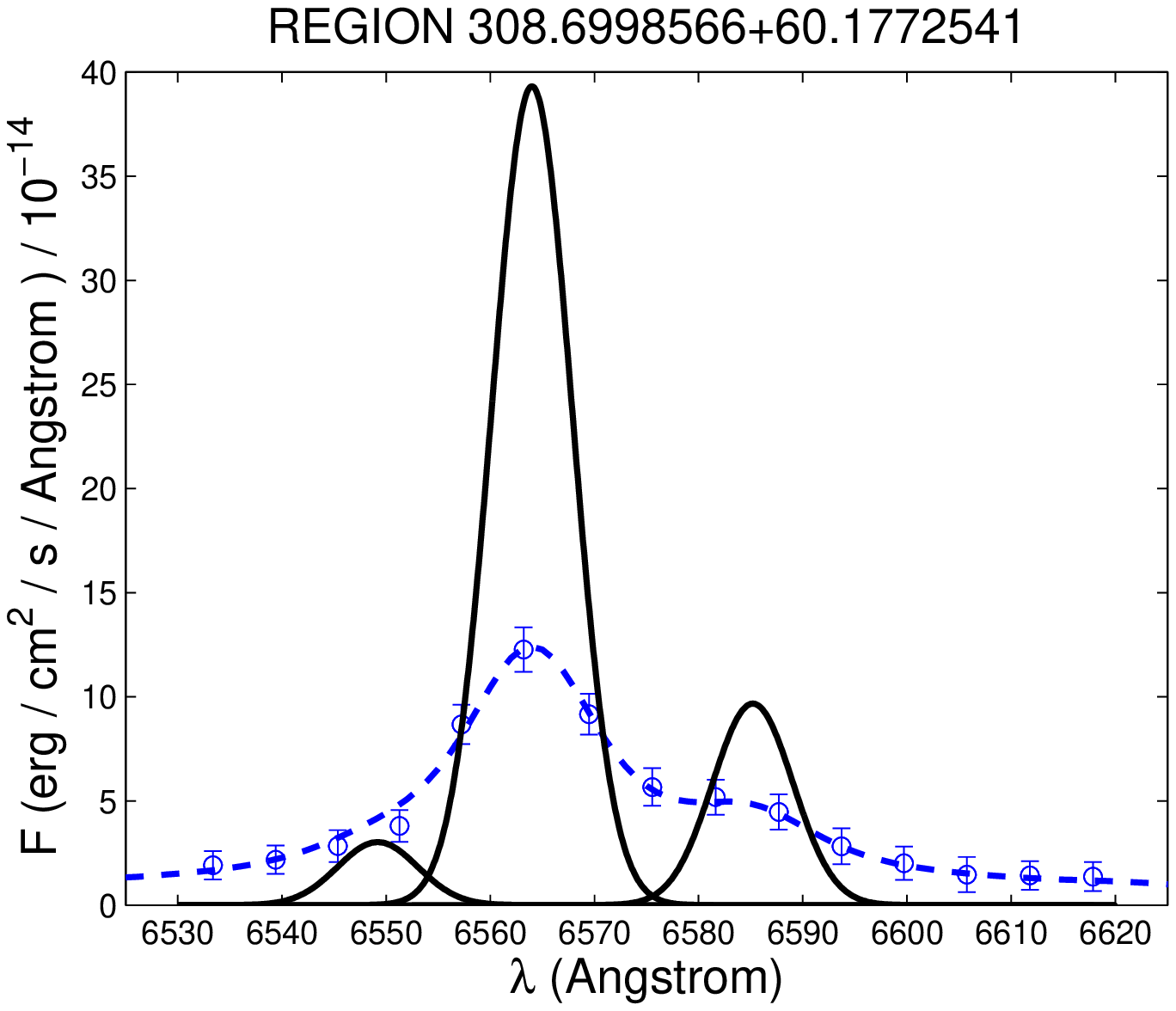} & \includegraphics[width=8cm]{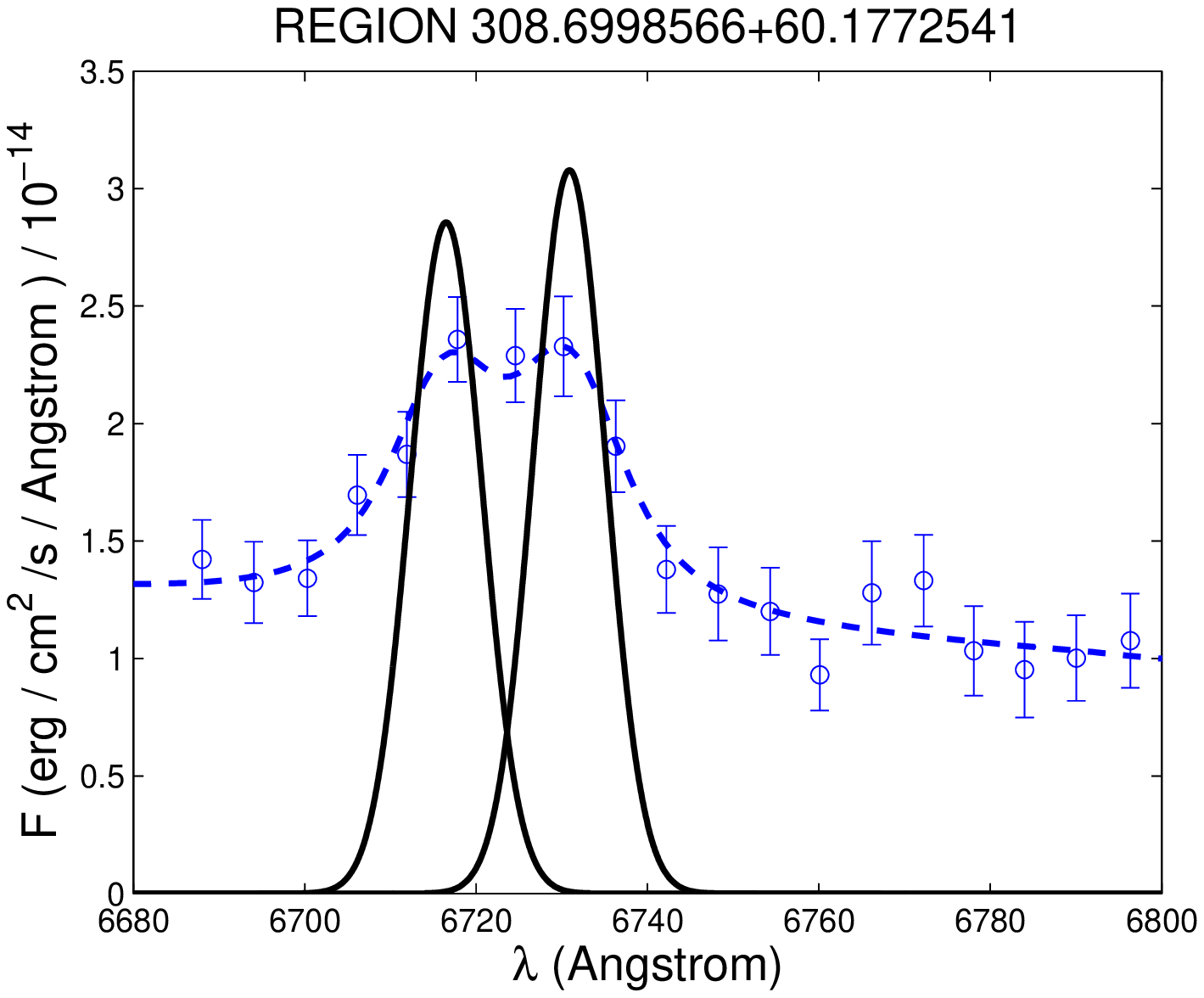}\\
 \end{tabular}
  \caption{Upper panel: composite image of NGC~6946 obtained by adding all the wavelengths for the H$\alpha$ scan. The circle indicates the position of region at RA: 308.6998566 degrees, DEC: +60.1772541 degrees. Lower panel: spectral energy distribution for H$\alpha$ + [NII] (left) and [SII] (right) zones. Open circles are the fluxes measured in each wavelength, the dashed line is the best fit for the data points, and the continuous lines are the deconvolved, deblended, and continuum subtracted gaussians for each emission line.}
\label{espec}
\end{figure*}

In Fig. \ref{espec} we show the resultant image of adding all the H$\alpha$ wavelengths (upper panel). A typical H~II region with enough signal to noise in H$\alpha$ and [SII] (RA: 308.6998566 degrees, DEC: +60.1772541 degrees) was selected as an example and its position is indicated by a circle in the image. In the lower panels is represented the spectral energy distribution of the selected region in the H$\alpha$+[NII] zone (left panel) and in the [SII] zone (right panel). The emission in each wavelength is indicated by the open circles, the dashed lines are the best fit to the data, and the continuous lines are the deconvolved components for each emission line.

\begin{figure}[!p]
 \plotone{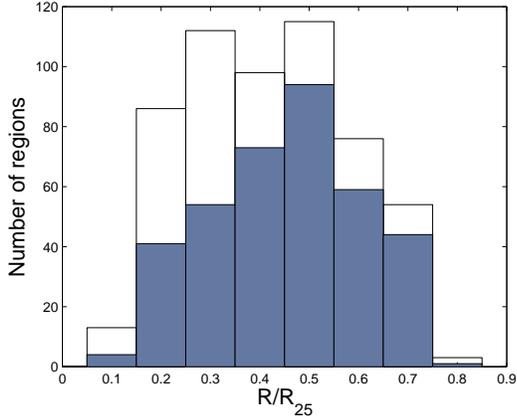}
 \caption{Number of detected regions as a function of the galactocentric radius, in bins of 0.1 in $R/R_{25}$. The numbers of regions detected in H$\alpha$ are the empty bars, the filled bars are the regions detected in [SII].}
 \label{his}
\end{figure}

The final catalogue has 557 H~II regions with H$\alpha$ emission detected and measured, with 370 of these H~II regions having measured fluxes in both of the [SII]$\lambda\lambda$6717,6731 lines. In Fig. \ref{his} is represented the numbers of detected regions in bins of 0.1 $R/R_{25}$ (with $R_{25}$=5.7$^\prime$, from de Vaucouleurs et al. 1991) for H$\alpha$ (empty bars) and for [SII] (filled bars).

\section{In situ and mean electron densities}

\subsection{In situ electron density}
The lines of R$_{[SII]}$=[SII]$\lambda$6717/[SII]$\lambda$6731 ratio are emitted by different levels with similar excitation energy, so this ratio is determined by the collisional strengths, and it is, therefore, sensitive to the electron density (Osterbrock, 1989).
To obtain the electron density from R$_{[SII]}$, we used the analytical solution given by McCall (1984) for a three-level atom:
\begin{equation}
{\rm R}_{[SII]}\simeq1.49\left[\frac{1+3.77\,x}{1+12.8\,x}\right]
\label{raz}
\end{equation}
where $x=10^{-4}n_et^{-1/2}$, $t=T/10000$, $n_e$ is the electron density in cm$^{-3}$, and $T$ is the electron temperature in K. If we assume that $t=1$ for all the regions, this gives us:
\begin{equation}
n_e=\frac{1.49\,{\rm R}_{[SII]}}{12.8\,{\rm R}_{[SII]}-5.6713}\times10^4
\label{den}
\end{equation}

According to Casta\~neda et al. (1992) or Zaritsky et al. (1994), the lower limit for reliably measured densities is 10\,cm$^{-3}$. However, due to the uncertainties in the fluxes for the [SII] lines, and the functional dependence of the density on the line ratio, we have selected a more conservative value of 50\,cm$^{-3}$ as a lower limit for the reliably measured density. With this limit, we were able to obtaint the electron density for 58 regions, well distributed within the disk of the galaxy.

\begin{figure}[!p]
\plotone{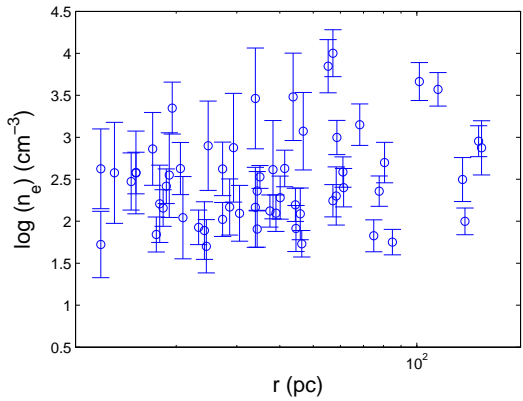}
\caption{Logarithm of the {\it in situ} electron density as a function of the equivalent radius for the measured H~II regions.}
\label{nerad}
\end{figure}

In Fig. \ref{nerad} we have represented the electron density as a function of the equivalent radius for the H~II regions. The values of the density range from $\sim$50\,cm$^{-3}$ (our imposed lower limit) up to $\sim$10000\,cm$^{-3}$, with a median value of 279\,cm$^{-3}$. There is no clear correlation betweeen these two parameters.

\subsection{Mean electron density}
The mean luminosity-weighted electron density $<n_e>$, is determined from the H$\alpha$ luminosity and the equivalent radius $r$ of each H~II region. Following Guti\'errez \& Beckman (2010) (thereafter GB10), this mean electron density can be expressed as:

\begin{equation}
<n_e>=1.5\times10^{-16}\sqrt{\frac{{\rm L}({\rm H}\alpha)}{r^3}}
\label{denj}
\end{equation}
Where L(H$\alpha$) is the luminosity in the H$\alpha$ line and $r$ is the equivalent radius of the H~II region. If L(H$\alpha$) is expressed in erg\,s$^{-1}$ and $r$ in pc, eq. \ref{denj} gives $<n_e>$ in cm$^{-3}$. The values found for $<n_e>$ lie in the range a few cm$^{-3}$ to a few tens cm$^{-3}$\ (see also Rozas et al., 1996, GB10).

\begin{figure*}[!p]
 \centering
 \begin{tabular}{c c}
  \includegraphics[width=8cm]{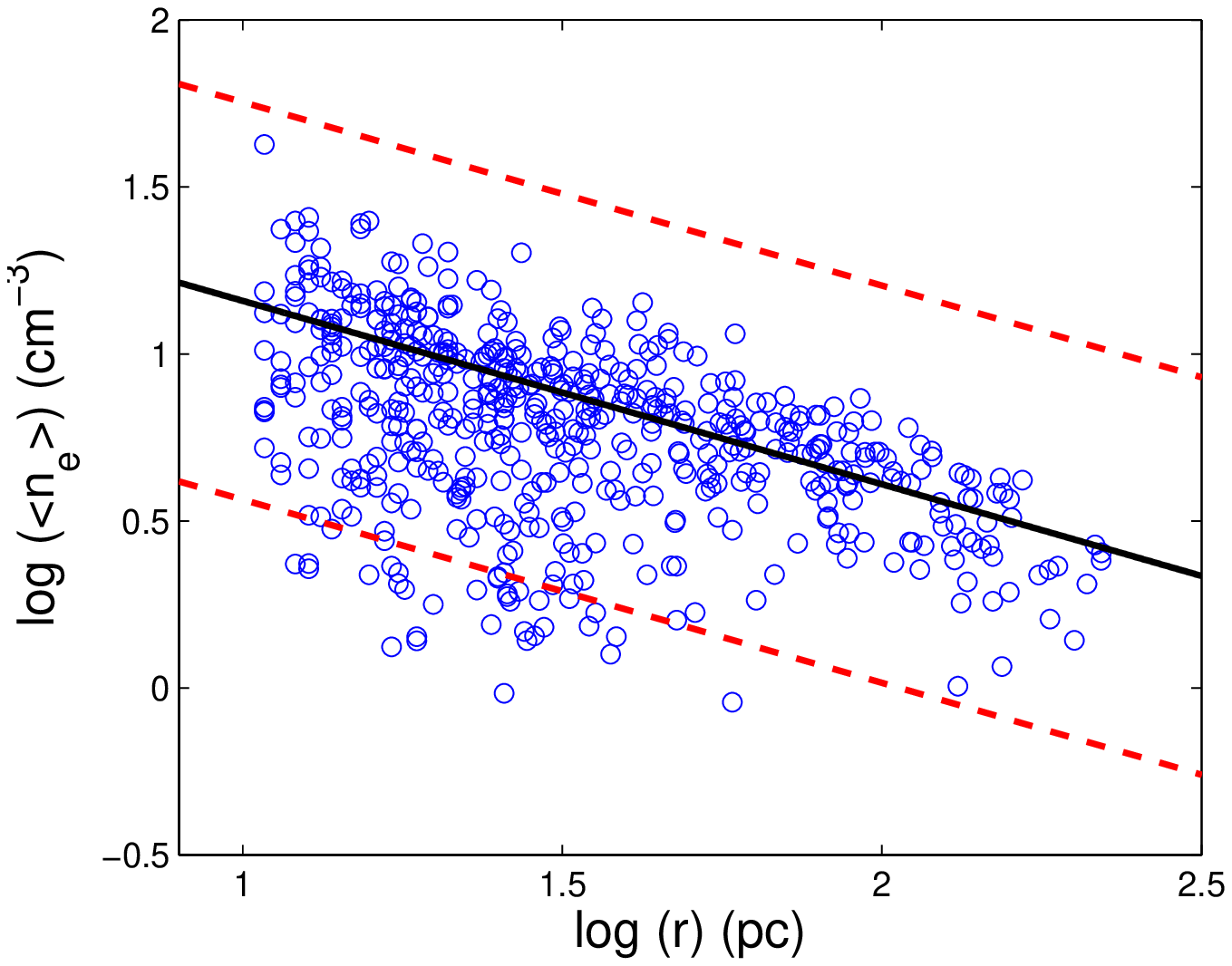} & \includegraphics[width=8cm]{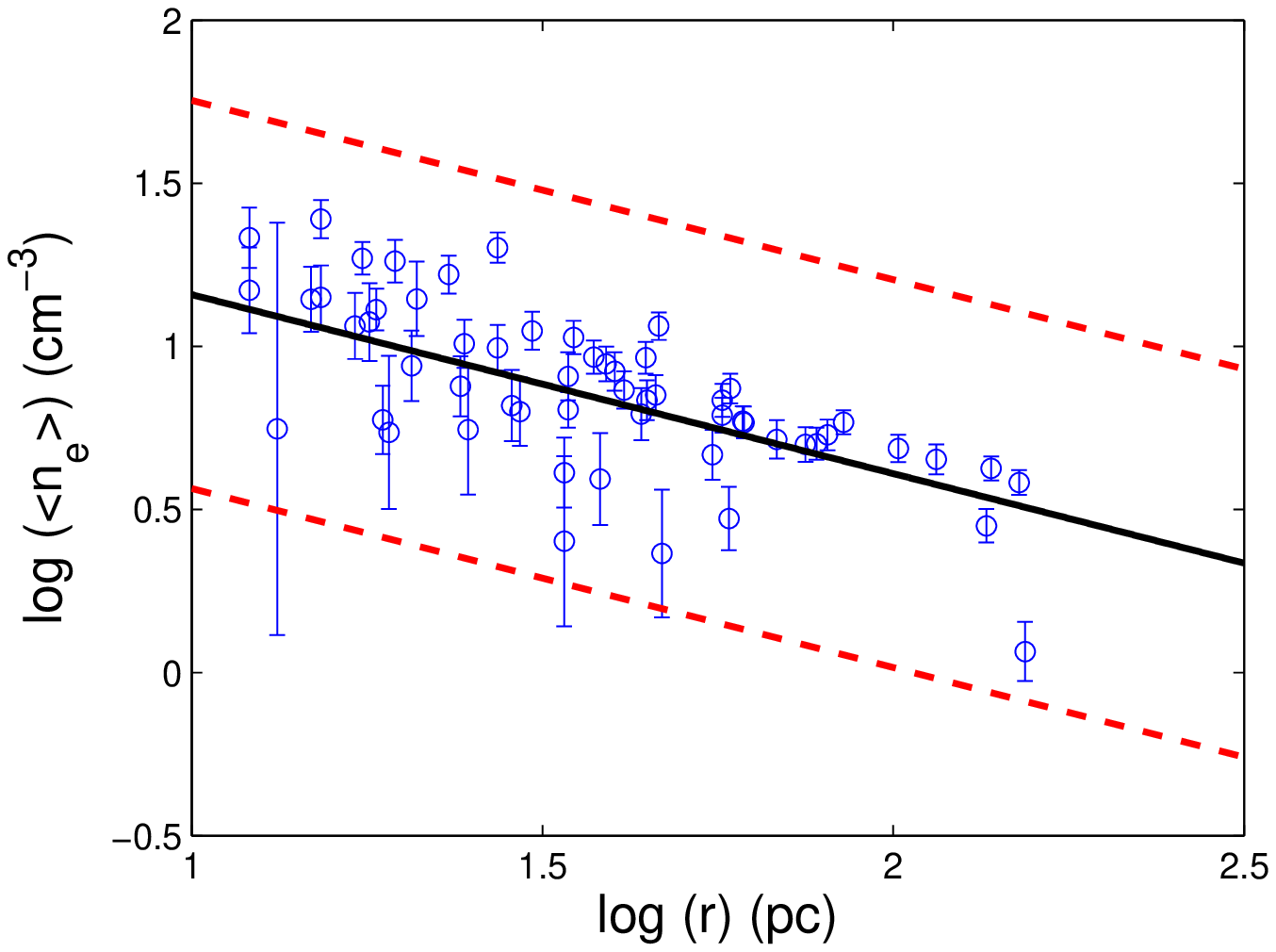}\\
 \end{tabular}
  \caption{Logarithm of the mean electron density, corrected by a factor which takes into account the galactocentric distance,  as a function of the logarithm of equivalent radius. The continuous line is the best fit for the data points. The dashed lines indicate the 2$\sigma$ confidence for the fit. In the left panel all the H~II regions with H$\alpha$ data are represented. In the right panel, only the regions with a determination of the {\it in situ} electron density are represented.}
\label{media}
\end{figure*}

In Fig. \ref{media} we represent the logarithm of the mean electron density as a function of the logarithm of the equivalent radius for all the H~II regions with H$\alpha$ data (left panel) and for the regions where we were able to determine the {\it in situ} electron density. We have corrected the mean electron density for the effect of the dependence on the galactocentric radius using the method proposed in GB10. A least squares fit to the data, gives the result

\begin{equation}
 <n_e>=\frac{51}{r^{0.55\pm0.05}},
\label{fit1}
\end{equation} 
which implies that the larger regions have lower mean electron densities. This result agrees with that presented in GB10, where they find that the relationship between the equivalent radius of the H~II regions and their mean electron density in two quite different galaxies: M51 and NGC~4449, follows the expression
\begin{equation}
<n_e>\sim\frac{1}{r^{0.5}}.
\label{fitg}
\end{equation}

\section{Filling factor}
We assume for an H~II region that the emission line ratios detected are weighted by density, as they are formed principally within the denser zones which occupy just a fraction of the total volume of the region, and that the rest of the H~II region is formed by a low density component. This component is sampled, at the same time as the high density clumps, when we derive mean electron density (Rozas et al. 1996, GB10). Following Osterbrock \& Flather (1959), we define the filling-factor ($ff$) of an H~II region as
\begin{equation}
ff=\left[\frac{<n_e>}{n_e}\right]^2.
\label{filing}
\end{equation}

\begin{figure}[!p]
\plotone{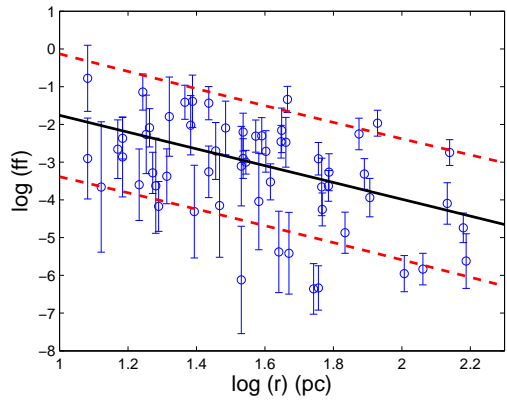}
\caption{Logarithm of the filling factor as a function of the logarithm of the equivalent radius of the regions. The continuous line is the best fit for the data points. The dashed lines indicate 66\% confidence level for the fit.}
\label{ffvsr}
\end{figure}

In Fig. \ref{ffvsr} we present the logarithm of $ff$ as a function of the logarithm of the equivalent radius of the H~II regions of NGC~6946. A clear anticorrelation exists between these parameters: larger H~II regions have lower values of the $ff$.  The correlation coefficent for this relationship is -0.3822, with a value of $p$=0.0004, where $p$ is the probability of getting a correlation as large as the observed value by random chance. In this case, $p$ is small enough (about 0.04\%\,) to conclude that the two quantities are related. A first degree polynomial (shown in the figure as a continuous line) was fitted to the data using the least squares method, and can be expressed as
\begin{equation}
ff=\frac{2.93}{r^{2.23}}.
\label{ff}
\end{equation}
This relationship indicates that, while the dense clumps within an H~II region maintain a density independent of the luminosity of the region (see Fig. \ref{nerad}), for larger H~II regions, there is a larger volume of low density gas filling the space between the denser zones. 

\section{Conclusions}
We have obtained reliable luminosities and radii for 557 H~II regions of NGC~6946 in H$\alpha$ and 370 of these H~II regions in both of the [SII]$\lambda\lambda$6717,6731 doublet lines.

From the doublet line ratio we could measure reliably {\it in situ} electron densities, $n_e$, for 59 H~II regions.
We found no correlation between the in situ electron density and the equivalent radius of the H~II regions. This seems to indicate that the denser zones where the bulk of the [SII] line emission is produced, show little variation in density over the full range of H~II region sizes and luminosities.

We also obtained the mean electron density $<n_e>$ for all the regions, using H$\alpha$ luminosities and the effective radii.
We found, as also found by GB10, a power law relationship with an exponent of -0.55, between $<n_e>$ and the equivalent radius of the H~II regions.

We then calculated the filling factor for the H~II regions of the sample. We found a power law relationship (with an exponent of -1.78) between the filling factor and the equivalent radius of the H~II regions, which indicates that larger H~II region are more porous. We can explain our results with the scenario in which the denser zones, with electron densities in the range 100 - 1000\,cm$^{-3}$, do not change their internal properties as we go from smaller, less luminous to large, more luminous regions, but they occupy a progressively smaller fraction of the volume. The tenous interclump gas, with density in the range 1-10\,cm$^{-3}$, occupies an increasing fraction of the volume the larger and more luminous the H~II regions are. We have produced the first quantitative estimate of this behaviour, which should open the way to more refined physical models of H~II region structure.

\acknowledgments

This work was supported by the Spanish Ministry of Economy and Competitiveness (MINECO) under grants AYA2011-29517-C03-01, AYA2007-67625-CO2-01, and AYA2010-17631, and project P3/86 of the Instituto de Astrof\'{\i}sica de Canarias. Based on observations made with the Gran Telescopio Canarias (GTC), instaled in the Spanish Observatorio del Roque de los Muchachos of the Instituto de Astrofísica de Canarias, in the island of La Palma. We wish to thank Dr. H\'ector O. Casta\~neda for their useful advice during the preparation of this study.

{\it Facilities:} \facility{GTC (OSIRIS)}.

\clearpage

\clearpage

\clearpage

\clearpage

\end{document}